# **Identifying Research Hotspots and Future Development Trends in *Current Psychology*: A Bibliometric Analysis of the Past Decade's Publications**

Shen Liu*, Yan Yang

Department of Psychology, School of Humanities and Social Sciences, Anhui Agricultural University, Hefei, Anhui, China

**Corresponding author**

Shen Liu

Department of Psychology, School of Humanities and Social Sciences, Anhui Agricultural University, No. 130 Changjiang Road(W), Shushan District, Hefei, Anhui, 230036, China

E-mail: liushen@ahau.edu.cn

**Acknowledgement**

This work was supported by the Outstanding Youth Program of Philosophy and Social Sciences in Anhui Province (2022AH030089) and the Starting Fund for Scientific Research of High-Level Talents at Anhui Agricultural University (rc432206).

**Author Contributions**

S.L. conceived and designed the study. S.L. and Y.Y. wrote and revised the main manuscript. S.L. and Y.Y. collected and analyzed the data. S.L. and Y.Y. prepared all figures and tables. All authors reviewed the manuscript.

**Declarations of interest**

None

**Data availability**

The datasets generated and analyzed during the current study are not publicly available. The datasets are available from the corresponding author on reasonable request when the aim is to verify the published results.

# Identifying Research Hotspots and Future Development Trends in *Current Psychology*: A Bibliometric Analysis of the Past Decade's Publications

**Abstract:** By conducting a bibliometric analysis on 4,869 publications in *Current Psychology* from 2013 to 2022, this paper examined the annual publications and annual citations, as well as the leading institutions, countries, and keywords. CiteSpace, VOSviewer and SCImago Graphica were utilized for visualization analysis. On one hand, this paper analyzed the academic influence of *Current Psychology* over the past decade. On the other hand, it explored the research hotspots and future development trends within the field of international psychology. The results revealed that the three main research areas covered in the publications of *Current Psychology* were: the psychological well-being of young people, the negative emotions of adults, and self-awareness and management. The latest research hotspots highlighted in the journal include negative emotions, personality, and mental health. The three main development trends of *Current Psychology* are: 1) exploring the personality psychology of both adolescents and adults, 2) promoting the interdisciplinary research to study social psychological issues through the use of diversified research methods, and 3) emphasizing the emotional psychology of individuals and their interaction with social reality, from a people-oriented perspective.

**Keywords:** *Current Psychology*; bibliometrics; information visualization; CiteSpace; VOSviewer

## 1. Introduction

Founded in 1881, *Current Psychology* is a well-established international journal that examines psychological topics from diverse perspectives. Additionally, the journal is dedicated to the rapid dissemination of cutting-edge, peer-reviewed psychological research. The journal welcomes important, rigorously empirical and theoretical works from across the major fields of psychology, including but not limited to: cognitive psychology and cognition, social psychology, clinical psychology, health psychology, developmental psychology, methodology and psychometrics, personality psychology, neuropsychology, human factors, and educational psychology. The latest impact factor of *Current Psychology* is 2.5. Contemporary psychology is developing rapidly, with an abundance of new research findings constantly emerging. *Current Psychology* is dedicated to identifying and highlighting the research hotspots and future development trends in international psychology. The journal systematically organizes and synthesizes three psychological research findings, exploring their implications of subsequent studies. In doing so, *Current Psychology* aims to predict the evolving trends and tendencies within the field of psychological research.

In recent decades, the rapid development of the social economy has led to an accelerated pace of life, giving rise to a growing number of psychological issues. Scholars are placing greater emphasis on journal-published research, as social issues within academia are frequently reflected in the works that are published. Correspondingly, the number of psychology publications has been steadily increasing in recent years. The sheer volume of research publications can make it challenging to fully grasp current research priorities and the overall state of the field, potentially leading to significant risks of overlooking fundamental issues and important areas for further study and practical advancements (Chan et al., 2017). In the past, the identification of research hotspots and development trends in a field was often based on the experiential judgments of individual experts, lacking systematic and demonstrable methods (Bussard, 2015; Maier, 2006; Peters et al., 2014). However, given the massive volume of publications in psychology, in order to elucidate the development trends of a specific field or to analyze the academic influence of a psychology journal and predict future trends in the discipline, we need to rely on objective indicators to be convincing. In the ongoing exploration and practice, using scientific measurement software to analyze this field is necessary to solve this problem (Chen, 2006). In recent years, some scholars have attempted to use bibliometrics to address this issue. They have started to use visualization analysis methods from the field of scientometrics to analyze psychology journals. Literature review is an effective way to gain an in-depth understanding of a research field. By systematically reviewing existing researches, we can understand the current status and development trends of the field in order to provide direction for future research (Zuo & Zhao, 2014). With the advancement of science and technology, many visualization tools have emerged in recent years, such as VOSviewer and CiteSpace. These tools support co-citation analysis and keyword co-occurrence analysis, which can help us conduct quantitative and objective analysis of related fields, revealing the quantitative relationships between various studies.

Some scholars have begun to use visualization analysis methods from the field of scientometrics to analyze psychology journals, and their findings are significant and highly valuable for reference and research. For example, Jia et al. (2019) explored the trends in psychological research based on the analysis of 3,814 publications in *Psychological Science* from 1990 to 2019. Tortosa-Pérez et al. (2021) conducted a bibliometric analysis of the literature published in *Anuario de Psychology Juridica* from 1991 to 2019 to explore the development trends of forensic psychology. Donthu et al. (2021) analyzed the publications of *Psychology and Marketing* from 1984 to 2020 to explore the association between psychology and marketing. Kataria et al. (2021) conducted a bibliometric analysis of publications in *Gender, Work and Organization* from 1994 to 2018 to explore the development status of the journal in its 25th year since its inception. Akturk (2022) explored the development of the *Journal of Computer Assisted Learning* since its founding 35 years ago, based on a bibliometric analysis of publications from 1985 to 2020. These studies all conducted visualization analysis on psychological journals, aimed at analyzing the developmental process of psychological research and looking ahead to future trends of psychology. However, no scholar has yet conducted a bibliometric analysis of publications in *Current Psychology* over a certain period, let alone predicting the hotspots and development trends of *Current Psychology* based on such an analysis. This paper also used bibliometric methods to analyze psychology journals, in order to clearly demonstrate the hotspots and thematic clusters. Specifically, this paper first combined CiteSpace, VOSviewer, and SCImago Graphica to analyze publications of *Current Psychology* from 2013 to 2022 in the Web of Science (WOS) database.

*Current Psychology* pays significant attention to social psychology, as well as human emotions, cognition, and affect, aiming to integrate current social situations and seek solutions to problems. The purpose of this paper included visualizing the quantity of annual publications, annual citations, the number of countries and institutions based on the publications of *Current Psychology*, evaluating their research performance, and revealing the geographical distribution characteristics of international cooperation and global institutions. Additionally, this paper provided an analysis of keywords, which may have potential guiding significance for researchers, especially those studying *Current Psychology* or the development of psychology as a whole, and provide a reference for future research.

## 2. Methods

### 2.1 Data source

The data for this paper was retrieved using an advanced search strategy in the WOS Core Collection on March 16$^{th}$, 2023. WOS is a high-quality digital database that has been widely used by researchers around the world and has become a common tool for retrieving and evaluating various types of publications (Thelwall, 2008). WOS covers a wide range of publications from different fields, including over 15,000 journals and more than 5 million classified publications in 251 categories and 150 research areas. In addition, WOS is a suitable database because it contains a set of data such as titles, authors, institutions, countries, abstracts, keywords, references, citation counts, and impact factors. The search terms entered in the publication title field were "Current Psychology", the language was set to English, the publication date ranged from January 1st, 2013, to December 31th, 2022, and the document types were limited to article and review article.

**2.2 Research tools and procedures**

Bibliometrics involves analyzing published information (such as books, journal articles, data sets, blogs) along with associated metadata (such as abstracts, keywords, citations) using statistical methods to depict relationships among publications (Broadus, 1987). Bibliometrics is a quantitative method used to assess and describe publications, aiding researchers in evaluating academic research within core areas (Rey-Martí et al., 2016; Small, 1973). Additionally, bibliometrics has been employed to trace the interconnections among academic journal citations. Nowadays, bibliometric analysis can analyze not only citations but also author collaboration networks, institutions, keyword co-occurrence networks, and journal distribution. The application of quantitative methods in bibliometrics analysis has introduced a more objective approach, enabling efficient analysis across a broader spectrum of publications. Nevertheless, traditional expert review remains irreplaceable as experts offer unique and insightful perspectives on the research field. With the aid of scientific statistical and bibliometric visualization software like CiteSpace and VOSviewer, data indicators such as annual publications, annual citations, authors, institutions, countries, and keywords for *Current Psychology* can be organized and analyzed. CiteSpace, renowned for its robust visualization and big data analysis capabilities, proves more comprehensive and apt for bibliometric analysis (Chen et al., 2015). It incorporates information visualization methods, bibliometrics, and data mining algorithms to visualize co-citation networks in scientific literature. By analyzing the annual publications, growth trends, and simultaneously occurring keyword clusters, it identifies keywords experiencing strong citation bursts over time and reveals trends and structures within knowledge fields. VOSviewer is another software utilized for constructing and visualizing bibliometric networks (van Eck & Waltman, 2010). It features user-friendly operation and presents a more concise diagram compared to CiteSpace. Although there are variations in how centrality and closeness are measured across nodes, the overall structure of the maps remains consistent and can be cross-verified (Zhang et al., 2011). These two software packages, CiteSpace and VOSviewer, collaborate to streamline the extensive workload in data analysis. They offer robust capabilities for knowledge network analysis, visually compelling graphical presentations, clear hierarchical structures, and straightforward interpretation of analysis results (Sabé, 2023). SCImago Graphica, developed by the SCImago team, utilizes the Google PageRank algorithm as a codeless visualization tool (Yan et al., 2020). Users can map data variables to different encoding channels and configure settings for each mapping, enabling the tool to swiftly generate interactive graphical displays for diverse data visualizations (Hirsch, 2005).

## 3. Development status of *Current Psychology*

### 3.1 Annual publications, annual citations, and publication trends

The trend of in publication numbers, encompassing quantity and growth rate, illustrates the relationship between publication count and time. It is a crucial indicator for measuring scholars' attention to a specific research field and reflects the overall progress of research within that domain (Li et al., 2023). Analyzing changes in the number of publications over a specific period helps to understand the development dynamics of research hotspots. The annual publications, citations per year, and publication trends of *Current Psychology* were depicted in Figure 1. The annual publications in *Current Psychology* have displayed a continuous upward trend. From 2013 to 2017, the number of publications remained stable, followed by a significant increase in 2018. The period spanning 2018 to 2021 saw rapid growth in publications, peaking in 2021, after which a declining trend began. The year 2018 marked a turning point for *Current Psychology*, with notable increases in both publication and citation rates. This expansion began in 2018 when *Current Psychology* significantly broadened its coverage to attract valuable, high-quality articles, thereby enhancing the journal's influence and international academic recognition. All impact factors were below 1.5 before 2018 and rose above 2 thereafter. Establish a process for special issues to attract outstanding papers. By aligning with contemporary needs, engaging a broad spectrum of scholars, publishing papers relevant to current societal issues, expediting the review process, and soliciting a greater volume of submissions. In 2021, *Current Psychology* reached another turning point, transitioning from continuous growth in publications to a sudden decline. This decline was attributed to the enduring impact of the COVID-19 pandemic on international politics, economics, and culture. Many countries and regions chose to implement lockdown measures, significantly impeding international scientific research cooperation and experimentation. As a consequence, the publications of *Current Psychology* started to decline. However, the regression results also indicated that the journal's annual publications have been growing rapidly since 2018, and this growth trend is projected to continue. This demonstrates that amidst the vigorous growth of international psychology research, *Current Psychology* is aligning with trends by increasing publications that hold theoretical and practical significance for psychological research.

In addition, data retrieved from WOS indicated that the average citation per article was only 4.6 times, significantly below the threshold for highly cited articles. The quality, research value, innovation, originality of articles, as well as the research field and object matter, all influence the number of citations an article receives. Therefore, *Current Psychology* should intensify its review efforts for submitted articles to ensure their originality and research value, aiming to improve the average citation frequency and thereby enhance the journal's recognition and impact. A special issue procedure in place for the journal to incorporate a wide range of scholarly perspectives, enhance relevance to contemporary issues, and improve the quality of published articles. The special issue undergoes rigorous evaluation with an emphasis on fairness in the review process. Enlist more reviewers, reduce review times, and solicit more excellent papers. Align with contemporary characteristics and develop special features that reflect current trends. The citations of *Current Psychology* demonstrated slow growth from 2013 to 2018, with a minor increase. However, starting from 2019, annual citations surpassed 1,000, reaching 9,603 citations by 2022, indicating significant growth. After fitting the journal's publication data, it was observed that the curve of the fitted publications displayed a notable upward trend, indicating an expectation for continued growth beyond 2022. Citation serves as an indicator of emerging trends (Chen et al., 2014). The rapidly growing number of citations indicates an increased practical application of papers in contemporary psychology, aligning with current trends. These results underscore the rising academic attention and industry recognition of the publications in *Current Psychology*, contributing to the continuous enhancement of the journal's overall recognition and influence.

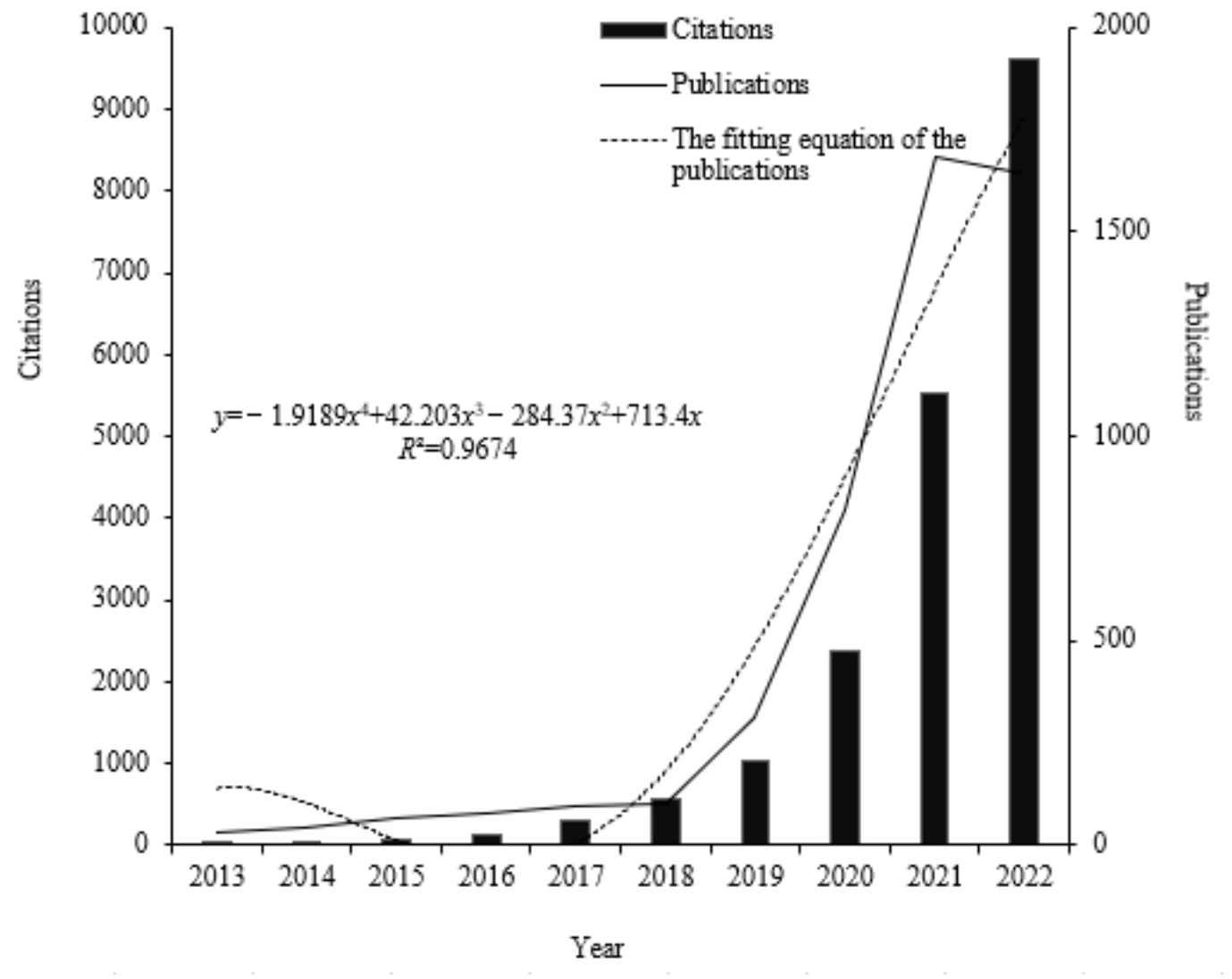

**Figure 1 Annual publications, annual citations, and publication trends of *Current Psychology***

### 3.2 Authors analysis

Table 1 displayed the top 20 authors by publications, highlighting that only the top 3 authors have accumulated over 10 publications each. Mark D. Griffiths ranked first with 12 publications, followed by Li Zhang with 11 publications, and Xuqun You with 10 publications. Overall, the differences variations among these authors were not substantial, and the distribution was relatively even. From the perspective of the author group, the contributors to *Current Psychology* exhibited a partially concentrated yet overall dispersed distribution. This indicates that the journal prioritizes articles quality over authorship focus, resulting in a dispersed distribution of article authors. The top-ranked author in terms of publications has only published 12 articles. Therefore, the sources of publications in *Current Psychology* are diverse, and the research scope is broad. The affiliations of these authors are also highly diverse, indicating that the journal welcomes submissions from a wide range of sources and adheres to principles of openness and impartiality, fostering sustainable and healthy development. Moreover, all of the top 20 authors in terms of publications had their first article published in 2017 or later, with a concentration of publications in 2021. This trend correlates with *Current Psychology*'s ongoing expansion in recent years, marked by rapid growth in publications.

**Table 1 Top 20 authors in *Current Psychology* by publications**

| Rank | Authors | Publications | Year | Current affiliations |
|---|---|---|---|---|
| 1 | Mark D. Griffiths | 12 | 2021 | University of Miami |
| 2 | Li Zhang | 11 | 2019 | Hunan Normal University |
| 3 | Xuqun You | 10 | 2017 | Shaanxi Normal University |
| 4 | Wei Wang | 9 | 2021 | Shaanxi Normal University |
| 5 | Marta Malesza | 8 | 2020 | Universität Potsdam |
| 6 | Alexandra Maftei | 8 | 2021 | Universitatea "Alexandru Ioan Cuza" |
| 7 | Claudia Ferreira | 8 | 2021 | Universidade de Coimbra |
| 8 | Ayse Altan-Atalay | 7 | 2020 | Koc University |
| 9 | Mengsi Xu | 7 | 2020 | Shaanxi Normal University |
| 10 | Yi Ding | 7 | 2021 | Changwon National University |
| 11 | Rude Liu | 6 | 2021 | Beijing Normal University |
| 12 | Scott E. Huebner | 6 | 2021 | University of South Carolina |
| 13 | Marcela Matos | 6 | 2021 | Universidade de Coimbra |
| 14 | Jon D. Elhai | 6 | 2021 | University of Toledo |
| 15 | Radoslaw Rogoza | 6 | 2019 | University of Social Sciences and Humanities in Warsaw |
| 16 | Stuart J. Mckelvie | 5 | 2017 | Bishop's University |
| 17 | Weijian Li | 5 | 2022 | Sichuan University |
| 18 | Ibrahim H. Acar | 5 | 2018 | Ozyegin University |
| 19 | Jing Wang | 5 | 2022 | Renmin University of China |
| 20 | Manpal Singh Bhogal | 5 | 2017 | University of Wolverhampton |

Table S1 in the Supplementary Materials presented the top 20 highly cited authors and their respective citation counts. In bibliometrics, the fundamental indicator for measuring the impact or intellectual influence of a paper is its citations. The influence of a scholar can be assessed based on the citations received by their publications. Overall, the citations of the authors were not concentrated and were relatively dispersed. Comparing with Table 1, several authors, such as Li Zhang and Xuqun You, appeared in both the top 20 lists for publications and citations. Authors with high publications and citation counts indicate that their work holds significant reference value and impact, aligning closely with the publication focus and development trends of *Current Psychology*.

We utilized CiteSpace to analyze the co-occurrence network of authors, generating a network diagram depicted in Figure S1 in the Supplementary Materials. The academic research ability of scholars can be reflected to a certain extent by their publications. The nodes in Figure S1 represented the frequency of authors' publications; authors with more publications have more nodes. In Figure S1, larger nodes indicate a higher number of publications. The lines between nodes in Figure S1 represented collaboration between two authors (Chen et al., 2021). According to Figure S1, the cluster appeared relatively dispersed and not prominently centralized, likely due to the large number of nodes. However, it is still evident that there was a certain level of collaboration between authors. Based on the network analysis of collaborative authors and the journal's publication history over the past decade, we can glean insights into the current research focus, direction of relevant authors, and their collaboration relationships.

**3.3 Countries and institutions analysis**

The 4,869 documents were processed using CiteSpace and Excel to consolidate institutions, merging secondary units into primary ones for analysis. The collected data was also converted for use in VOSviewer. The threshold was set at 1, indicating that institutions with a frequency greater than 1 could be visualized. In Figure S2 of the Supplementary Materials, it is evident that 12 out of the top 20 high-productivity institutions are based in China, highlighting a significant contribution of Chinese scholars to publications in *Current Psychology*. The top three institutions in terms of publications were Beijing Normal University, followed by Southwest University, and the third was Renmin University of China. Table S2 in the Supplementary Materials displayed the top 20 high- productivity institutions in *Current Psychology*, with12 of them located in China. It is evident that Chinese scholars have made significant contributions to publications in *Current Psychology*. Beijing Normal University led with the highest number of publications, totaling 176 articles, accounting for 3.62% of the total publications. Southwest University had 70 publications, accounting for 1.44% of the total, while Renmin University of China had 61 publications, accounting for 1.25% of the total. The top-ranked Beijing Normal University had significantly more publications than other institutions, highlighting its strong research capacity in the field of psychology. Beijing Normal University has a long history in psychology, and its Faculty of Psychology is recognized as a world-class institution in the field. It is the sole national key discipline unit for psychology, and in the Ministry of Education of People's Republic of China's discipline rankings, its psychology department has consistently held the top position in China. Both "Psychiatry and Psychology" and "Neuroscience and Behavioral Sciences" have entered the top 1% of the ESI rankings. The publications of institutions ranked lower did not show significant differences among them. The Faculty of Psychology at Southwest University hosts a national key discipline in basic psychology, operates a postdoctoral research station in psychology, and offers a doctoral degree program in psychology. It also offers authorized Master's degrees programs in applied psychology and mental health education within the field. The Institute of Psychology at Renmin University of China is equipped with advanced experimental facilities that enables extensive studies in neuroscience and psychology. This infrastructure supports in-depth research into the complex dynamics of human thinking, behavior, and social behaviors. It is precisely because these institutions focus on psychology research and pay attention to current topics that they achieve outstanding academic results, representing cutting-edge developments in domestic psychology and continually advancing the field in China. The top 20 institutions, excluding the 12 from China, were all from Europe or North America, regions characterized by relatively developed economies. Scientific research and economic strength are mutually reinforcing: advances in research drive economic growth, while economic development provides essential support for research and development activities, necessitating substantial research funding. Psychology has a long-standing history in Europe and continues to evolve. Diversified research orientations, innovative paradigms, cross-disciplinary integration, expanded fields, and emerging trends, methods, and theories have collectively fostered a rich and diverse developmental landscape in European psychology. In contrast, psychology in the United States shows some deviations from traditional approaches, marked by emerging trends emphasizing applications and the development of applied psychology.

Analysis of countries and institutions in the knowledge graph can reflect the distribution of papers and collaborations among different countries and institutions. As shown in Figure S3 of the Supplementary Materials, the size of the nodes reflects the quantity, and the link value between the nodes reflects the strength of collaboration (Xia & Zhong, 2021). By utilizing CiteSpace's country collaboration analysis function, we can assess the publications and their impact across different countries. The country distribution of literature can reflect the geographical distribution of active authors worldwide and each country's contribution. The Country Collaboration Analysis Network enables the analysis of article volumes originating from different countries, providing insights into widespread journal publications and robust inter-country connections. The greater the number of countries involved, the greater the impact of the journal can be demonstrated. Combined with Table S2, it is evident that the top three institutions in terms of publications were all located in China. Table S3 showed that the top three countries in terms of publications were People’s Republic of China, the United States, and the Republic of Türkiye, accounting for 27.83%, 19.16%, and 8.61%, respectively. There were significant differences in publications among scholars from various countries, with Chinese scholars contributing significantly to publications in *Current Psychology*. According to Table S3 in the Supplementary Materials, countries with higher publication outputs demonstrated stronger economic and overall capabilities. Strong national capabilities can support the continuous development of a scientific research within a country, and substantial investments in research funding can further enhance scientific output. The link strength reflects the intensity of collaboration (Wang et al., 2022). As shown in Figure S3, collaboration among different countries were extensive and closely intertwined, forming tight and complex collaboration chains. Scientific research thrives on cooperation, which fosters long-term and sustainable development.

According to the geographic visualization in Figure 2, it is evident that there were three main clusters of countries with high publication output: East Asia, North America, and Europe. These regions exhibit high levels of comprehensive economic development, providing robust support for scientific research. Among them, the People’s Republic of China leads in East Asia, while the United States i dominates in North America. In Europe, the distribution was more dispersed, with multiple countries showing comparable publication levels and less distinct clustering.

China focuses on keywords such as "depression", "COVID-19" and "adolescents", indicating its primary concerned with the mental health of specific demographic groups during critical periods. The United States pays closer attention to keywords such as "social support" and "self-efficacy", indicating a focus on individual development. The findings of several studies suggest China's focus on psychological impacts amidst COVID-19. Publications from China have shown a significant growth trend, surpassing other regions. However, there remains an imbalance in publication levels across various regions, with many having minimal or no contributions, highlighting the necessity for enhanced regional cooperation.

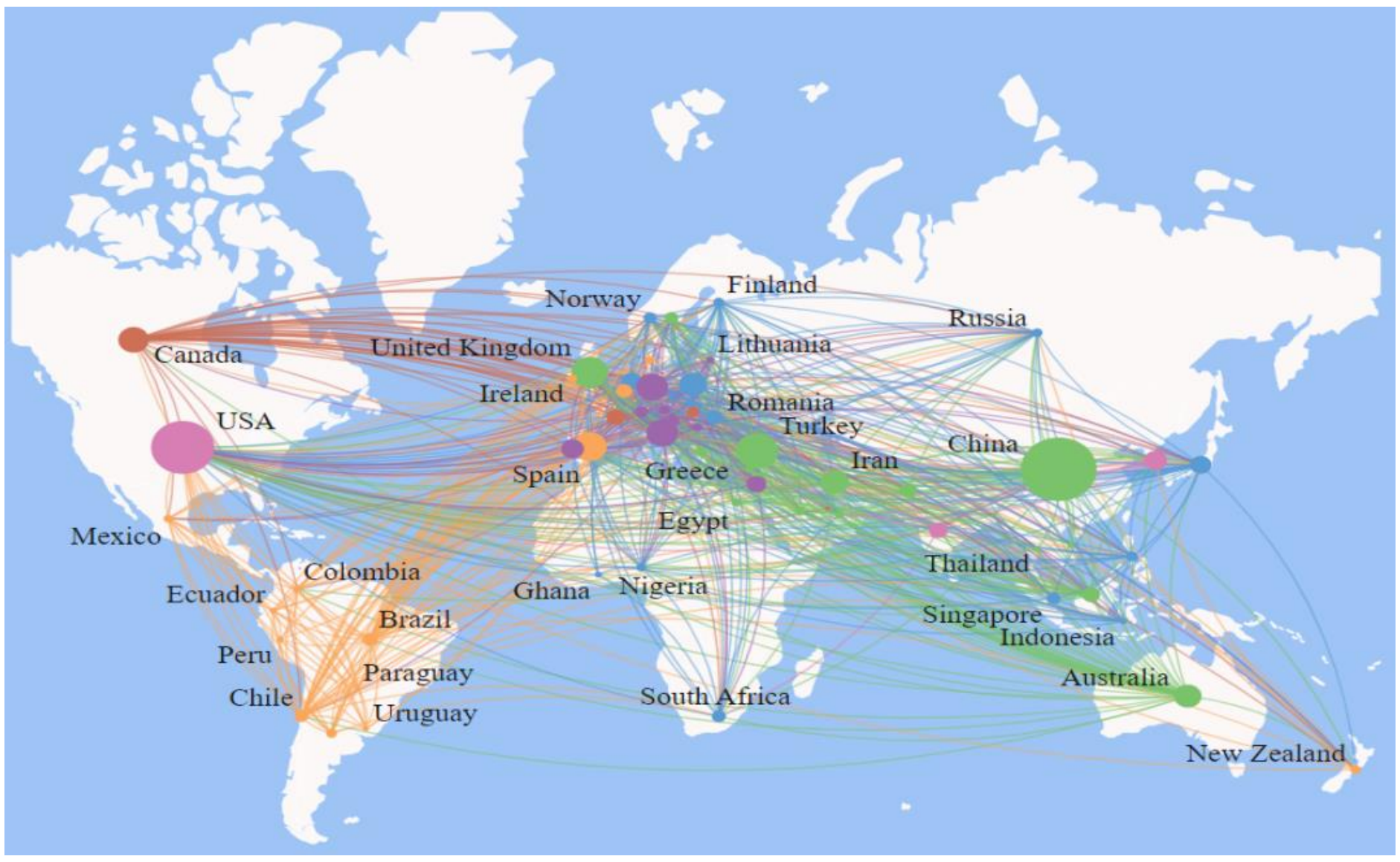

**Figure 2. Geographic visualization view**

### 3.4 Highly cited references analysis

Citation is an important indicator for measuring the level and academic influence of a journal. The number of times an article is cited reflects the quality of the articles published in the journal and the research level of scholars and their affiliated institutions. Citation is an index to measure the quality of papers and is often referred to as citation frequency. The higher the citation frequency, the greater the academic value. Citation analysis is an important method for assessing the impact of an article on scientific progress and evaluating the impact factor of scientific journals (Moed, 2009). Citation analysis is a valuable bibliometric method introduced in 1987, widely applied across various fields for assessing the impact of articles and journals, proving significant for authors and publications alike. For authors, citation analysis serves not only to identify crucial research advancements but also provide valuable perspectives on the historical development of their academic interests. For journals, data from citation analysis may attract manuscripts with higher citation potential. Additionally, citation analysis can help researchers to enhance their work to some extent (Xue et al., 2018). Highly cited documents are papers that are frequently referenced within a field of study, often regarded as foundational and key references for research. The frequency of reference use represents its influence, while co-citation reveals research topics and development background (Zhao, 2022). Table 2 listed the top 10 cited references, covering studies on depression, happiness, arousal, dominance, psychological strengths and resilience, compensatory behavior, attention, practice, perceived stress, and self-fulfilling prophecies.

**Table 2. Top 10 highly cited articles in *Current Psychology* from 2013 to 2022**

| Rank | References | Citations from WOS | Citations from google scholar |
|---|---|---|---|
| 1 | Karim, J., Weisz, R., Bibi, Z., & Rehman, S. R. (2015). Validation of the eight-item center for epidemiologic studies depression scale (CES-D) among older adults. *Current Psychology, 34*, 681–692. | 162 | 234 |
| 2 | Bakker, I., van der Voordt, T., Vink, P., & de Boon, J. (2014). Pleasure, arousal, dominance: Mehrabian and Russell revisited. *Current Psychology, 33*, 405–421. | 147 | 367 |
| 3 | Yıldırım, M., & Arslan, G. (2022). Exploring the associations between resilience, dispositional hope, preventive behaviours, subjective well-being, and psychological health among adults during early stage of COVID-19. *Current Psychology, 41*, 5712–5722. | 138 | 209 |
| 4 | Datu, J. A. D., Valdez, J. P. M., & King, R. B. (2016). Perseverance counts but consistency does not! Validating the short grit scale in a collectivist setting. *Current Psychology,* 35, 121–113. | 124 | 334 |
| 5 | Frayn, M., & Knäuper, B. (2018). Emotional eating and weight in adults: A review. *Current Psychology, 37*, 924–933. | 117 | 248 |
| 6 | Grzesiak-Feldman, M. (2013). The effect of high-anxiety situations on conspiracy thinking. *Current Psychology, 32*, 100–118. | 115 | 210 |
| 7 | Goretzko, D., Pham, T. T. H., & Bühner, M. (2021). Exploratory factor analysis: Current use, methodological developments and recommendations for good practice. *Current Psychology*, *40*, 3510–3521. | 114 | 216 |
| 8 | Mondo, M., Sechi, C., & Cabras, C. (2021). Psychometric evaluation of three versions of the Italian perceived stress scale. *Current Psychology, 40*, 1884–1892. | 100 | 124 |
| 9 | Alper, S., Bayrak, F., & Yilmaz, O. (2021). Psychological correlates of COVID-19 conspiracy beliefs and preventive measures: Evidence from Turkey. *Current Psychology*, *40*(11), 5708–5717. | 84 | 155 |
| 10 | Li, M. J., Wang, Z. H., Gao, J., & You, X. Q. (2017). Proactive personality and job satisfaction: The mediating effects of self-efficacy and work engagement in teachers. *Current Psychology, 36*, 48–55. | 82 | 227 |

CiteSpace was used with default settings to cluster co-cited references and select label clusters using indexing terms. The cited references form the knowledge base of the research. The frequency of references use represents its influence, while co-citation reveals research topics and development background. Based on the references cluster analysis in Figure S5 of the Supplementary Materials, several hotspots and trends relevant to *Current Psychology* were identified, characterized by high citation rates and significant reference value. Figure S6 in the Supplementary Materials displayed the top 20 cited references with the strongest citation bursts in publications of *Current Psychology* over the past decade.

### 3.5 Highly cited keywords analysis

Table 3 listed the top 20 keyword frequency rankings for *Current Psychology*, as along with their first appearance year and centrality. Centrality is an index used to determine the importance of nodes within a network. The higher the centrality, the more important the node is in the network. In terms of centrality, "performance", "personality", "mental health", and "model" scored higher, indicating their significant role in the publication network of *Current Psychology*. As shown in Table 3, emotions and mental states were central to *Current Psychology*, with "depression" appearing 373 times, "model" 318 times, "personality traits" 304 times, and "psychological health" 302 times. In addition, research on adolescents, social support, and gender differences is also highly valued. Table 3 clearly illustrates that publications in *Current Psychology* focus heavily on human psychology, perception, and cognition, establishing it as a research hotspot for mental and physical health, aimed at enhancing human development through psychological insights.

**Table 3. Keywords in *Current Psychology* from 2013 to 2022**

| Rank | Keywords | Counts | Year | Centrality |
|---|---|---|---|---|
| 1 | depression | 373 | 2015 | 0.02 |
| 2 | model | 318 | 2013 | 0.04 |
| 3 | personality | 304 | 2014 | 0.04 |
| 4 | mental health | 302 | 2014 | 0.04 |
| 5 | validation | 297 | 2015 | 0.02 |
| 6 | stress | 288 | 2014 | 0.02 |
| 7 | behavior | 268 | 2014 | 0.05 |
| 8 | performance | 260 | 2013 | 0.07 |
| 9 | anxiety | 257 | 2013 | 0.01 |
| 10 | health | 256 | 2015 | 0.03 |
| 11 | impact | 222 | 2015 | 0.02 |
| 12 | adolescents | 211 | 2013 | 0.03 |
| 13 | scale | 209 | 2014 | 0.05 |
| 14 | children | 201 | 2014 | 0.03 |
| 15 | symptoms | 190 | 2017 | 0.02 |
| 16 | satisfaction | 182 | 2014 | 0.02 |
| 17 | validity | 179 | 2015 | 0.03 |
| 18 | social support | 164 | 2014 | 0.03 |
| 19 | self | 162 | 2015 | 0.02 |
| 20 | gender differences | 153 | 2013 | 0.03 |

The burstness of a keyword signifies its importance as a research hotspot within a specific period , offering insights into trending research areas. It aids in observing the research trends that garner attention during that time. The burstness of keywords was calculated using the "Burstness" function, and the results were displayed in Figure S7. The blue line segments indicated time intervals, while the red line segments represented keyword bursts observed during the analyzed period. Gender differences, cognition, and performance have emerged as the primary research directions in *Current Psychology*. According to the ranking, the burst intensity of *Current Psychology* keywords in the generated WOS database (as depicted in Figure S7 in the Supplementary Materials) could be roughly divided into two stages. In the first stage, the primary burst word from 2013 to 2014 was "gender differences", making the earliest high-frequency burst word with an intensity of 8.84. This indicated that during 2013–2014, *Current Psychology* placed significant emphasis on issues related to gender differences. Adolescents' attitudes, perceptions, and performance also served as the foundation for research in this field. In the second stage, from 2015 to 2019, the main emerging keywords were "self-personality", "values", and "satisfaction". This shift reflects the growing societal focus on human physical and mental health, which is clearly evident in the publications of *Current Psychology*.

## 4. Hotspot analysis and trends prediction of *Current Psychology*

Keywords cluster analysis can unveil the current status and research frontiers of specific topics. This method employs algorithm to identify homogeneous units within the data. Due to the close relationship between keywords and the core of the literature, analyzing similar keywords can assist in identifying the research focal points. This study employed cluster analysis on high-frequency keywords, incorporating publication times into its analysis. This approach not only highlights research hotspots through clustering but also clarifies research directions and topics. The timeline view focuses on the evolution of research within each cluster, reflecting the developmental trends across various research directions. Based on the analysis of keywords co-occurrence, this paper conducted cluster analysis on high-frequency keywords, incorporating

publication times into its analysis. This approach not only highlights research topics through clustering but also provides clearer insights into research directions and topics compared to keywords co-occurrence analysis. The timeline view displays clustering information on the *y*-axis and publication year on the *x*-axis. By observing this view, analysts can examine the starting year and duration of clusters.

In Figure S8 of the Supplementary Materials, each "#" represented a node, with the size of each node indicating the frequency of the corresponding phrase. The lines connecting the nodes indicated that the two connected nodes appeared in the same article, with the thickness of the line proportional to the strength of their connection. The modularity $Q$=0.39 (>0.30) and weighted average silhouette $S$=0.64 (>0.50), indicate that the clustering exhibits strong performance and a coherent structure, which is compelling (Li et al., 2023).

Keywords are pivotal concepts in an article, typically appearing in 3–5, providing one of the simplest ways to grasp the paper's topic. Therefore, analyzing keywords can uncover the research hotspots within the field. As highly summarized representations of research content, the co-occurrence network and keyword frequency provide researchers with a clear view of current research hotspots in psychology. This highlights the research questions emphasized during a specific period and aids in identifying future research trends. Analyzing the keywords in literature helps to grasp the research hotspots, knowledge structure, and developmental direction of the field. Keywords with high frequency and centrality often signify the research hotspots and trends in the field. This paper employed VOSviewer for cluster analysis on the keywords of 4,869 publications in *Current Psychology*. The results were depicted in Figure 3, with detailed findings presented in Table 4. The keyword cluster network can be visualized using VOSviewer, which can highlight hotspots (Chen, 2012). From Figure 3, the keyword co-occurrence network reveals several clusters, each with representative themes: The yellow cluster, centered around "depression", explores the impact of negative emotions on individuals. The blue cluster, focused on "performance," examines aspects of job performance. Lastly, the purple cluster, highlighting "mental health", investigates mental health within various social contexts. Key terms such as "Depression", "personality", "mental health" and "performance" emerge as highly cited within their respective clusters.

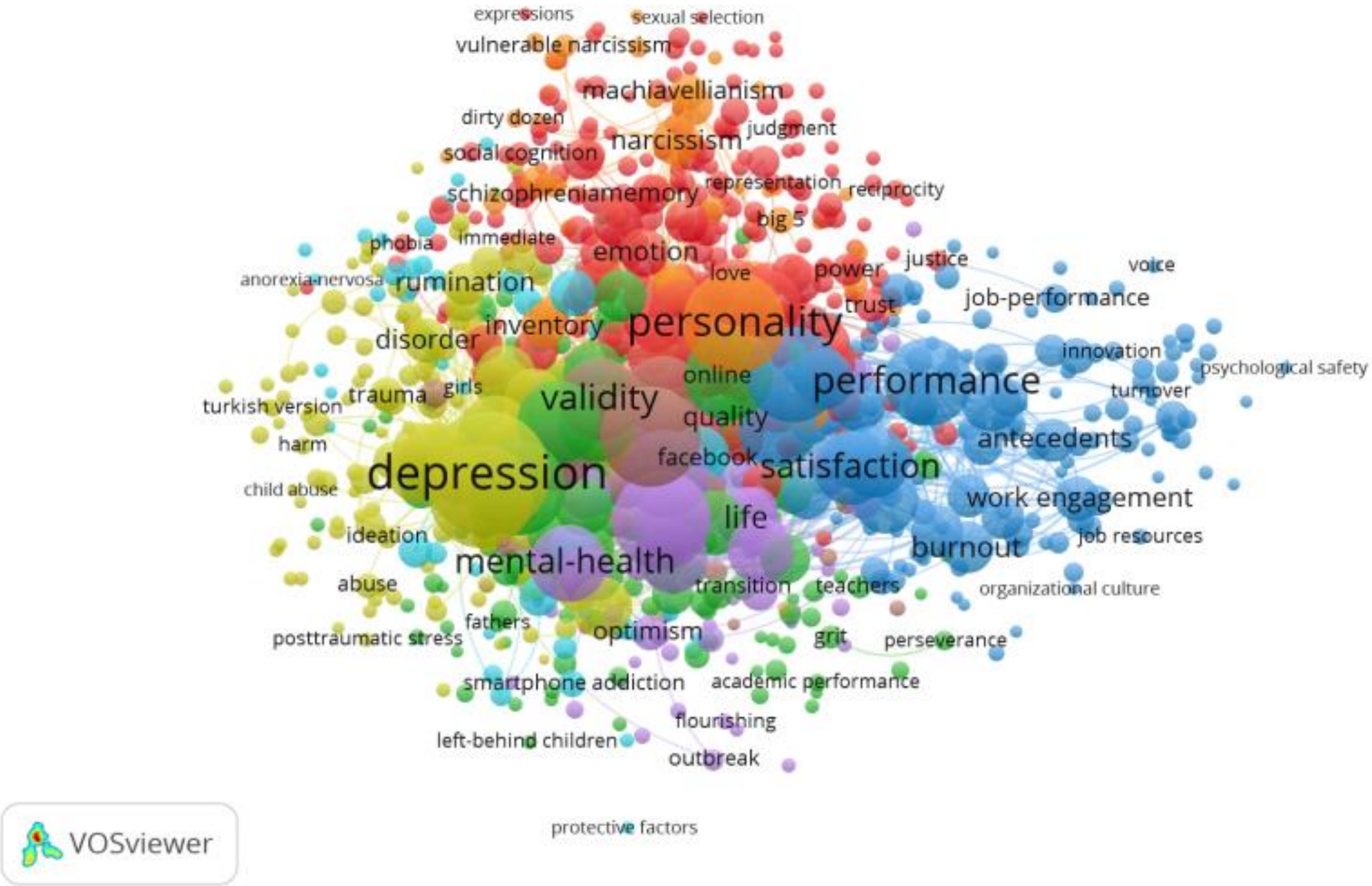

**Figure 3. Thematic keywords cluster network via VOSviewer**

To clearly and concisely observe clustering hotspots and gain an intuitive understanding of the development process of *Current Psychology*, CiteSpace was utilized to generate a timeline graph depicting the journal's research hotspots. The timeline graph can display the associations between different clusters and the evolution of keywords within these clusters over time. The timeline from 2013 to 2022 was divided into a series of time periods by CiteSpace to generate a co-citation network. Each cluster was displayed on a horizontal line with a corresponding label on the right. The timeline was positioned at the top to indicate the publication dates. These lines represent co-citation links between nodes, with colors distinguishing between different clusters. The timeline view can reflect the evolving trends of the research field over time, highlighting emerging and pivotal research directions. The first 14 frequently cited articles were selected for further analysis and clustered into 14 categories (numbered 0 to 13), where lower numbers indicate more keywords in each cluster. Each cluster comprises multiple closely related phrases, as depicted in Figure 4.

Cluster #0 focused on "work engagement", Cluster #1 on "working memory", Cluster #2 on "self-determination theory", Cluster #3 on "young adults", Cluster #4 on "dark triad", Cluster #5 on "parenting stress", Cluster #6 on "mental health", Cluster #7 on "life satisfaction", Cluster #8 on "academic performance", Cluster #9 on "suicidal ideation", Cluster #10 on "body image", Cluster

#11 on "emotion regulation", Cluster #12 on "mild cognitive impairment", and Cluster #13 on "virtual reality". As shown in Figure 4, the largest clusters were Cluster #0 "work engagement" and Cluster #1 "working memory", which emerged early and persisted over time. Cluster #3 "young adult" showed earlier but saw a decline in mentions from 2015 to 2019, while Clusters #9 "suicidal ideation", #10 "body image", and #11 "emotion regulation" emerged relatively later. These results highlight that prevalent themes in *Current Psychology* have focused on topics such as work engagement, working memory, young adults, and mental health. Work engagement and working memory have consistently garnered research attention over the past decade. In contrast, suicidal ideation, body image, and emotion regulation have seen increased interest in recent years. Moreover, young adults are experiencing a resurgence in research attention after a period of relative neglect.

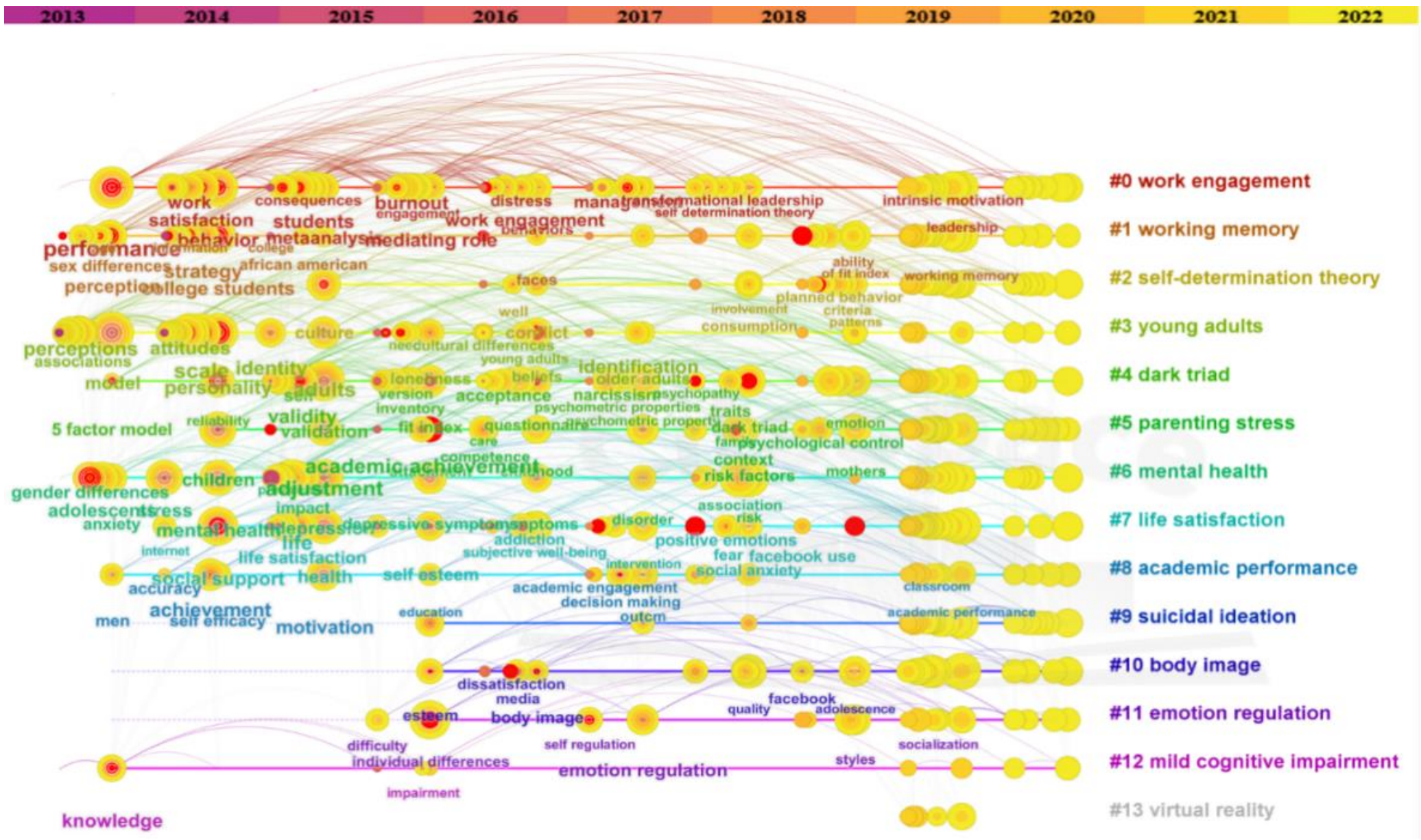

**Figure 4. Timeline view of co-occurrence of keywords**

Based on the LLR algorithm, Table 4 was generated. Combined with Figure 4, the keywords were predominantly categorized into three groups. The first group reflected themes such as early adulthood engagement in work, job satisfaction, early adulthood, and executive function, illustrating adults' dedication to their careers. The second category highlighted various societal pressures faced by adults, including parenting stress, depression, social identity, the COVID-19 virus, and psychological distress. The third category highlighted extreme negative emotions experienced by individuals, including suicidal ideation, social comparison, sadness, and emotion regulation suppression, which exert significant social pressure. Therefore, *Current Psychology* focuses extensively on the cognitive aspects and negative emotions of individuals, reflecting the journal's thematic emphasis.

**Table 4. LLR algorithm of keywords**

| Cluster ID | Size | Silhouette | mean (Year) | Label (LLR) |
|---|---|---|---|---|
| 0 | 70 | 0.778 | 2017 | **work engagement (115.56, 1.0E–-4)**; job satisfaction (79.35, 1.0E–4); emotional exhaustion (54.12, 1.0E–4); core self-evaluations (31.87, 1.0E–4); organizational commitment (30.73, 1.0E–4) |
| 1 | 70 | 0.654 | 2018 | **working memory (32.94, 1.0E–4)**; latent profile analysis (26.54, 1.0E–4); attractiveness (23.54, 1.0E–4); social cognition (22.05, 1.0E–4); executive functions (16.46, 1.0E–4) |
| 2 | 70 | 0.661 | 2020 | **self-determination theory (44.2, 1.0E–4)**; basic psychological needs (24.66, 1.0E–4); perceived risk (23.57, 1.0E–4); construal level (17.67, 1.0E–4); emerging adults (16.15, 1.0E–4) |
| 3 | 69 | 0.668 | 2016 | **young adults (30.57, 1.0E–4)**; sexual orientation (26.88, 1.0E–4); social identity (26.29, 1.0E–4); older adults (25.18, 1.0E–4); stigma (22.28, 1.0E–4) |
| 4 | 62 | 0.721 | 2018 | **dark triad (83.74, 1.0E–4)**; measurement invariance (61.9, 1.0E–4); confirmatory factor analysis (53.86, 1.0E–4); psychometric properties (50.26, 1.0E–4); personality (37.59, 1.0E–4) |
| 5 | 60 | 0.673 | 2020 | **parenting stress (31.38, 1.0E–4)**; parenting styles (25.3, 1.0E–4); covid-19 pandemic (21.87, 1.0E–4); peer victimization (16.06, 1.0E–4); early adolescents (15.68, 1.0E–4) |
| 6 | 57 | 0.754 | 2018 | **mental health (75.37, 1.0E–4)**; depression (58.62, 1.0E–4); psychological distress (32.41, 1.0E–4); anxiety (27.2, 1.0E–4); covid-19 (27.14, 1.0E–4) |
| 7 | 51 | 0.717 | 2018 | **life satisfaction (47.97, 1.0E–4)**; subjective well-being (45.96, 1.0E–4); social support (40.5, 1.0E–4); mindfulness (28.54, 1.0E–4); positive affect (28.47, 1.0E–4) |
| 8 | 39 | 0.745 | 2019 | **academic performance (27.84, 1.0E–4)**; academic engagement (24.14, 1.0E–4); student engagement (22.81, 1.0E–4); grit (16.59, 1.0E–4); motivation (14.24, 0.001) |
| 9 | 38 | 0.791 | 2021 | **suicidal ideation (29.99, 1.0E–4)**; psychosis (18.14, 1.0E–4); post-traumatic stress disorder (15.7, 1.0E–4); sleep quality (13.73, 0.001); abusive supervision (13.73, 0.001) |
| 10 | 26 | 0.823 | 2019 | **body image (30.88, 1.0E–4)**; social media (28.49, 1.0E–4); body dissatisfaction (21.55, 1.0E–4); social comparison (17.81, 1.0E–4); repetitive negative thinking (17.81, 1.0E–4) |
| 11 | 25 | 0.886 | 2020 | **emotion regulation (92.12, 1.0E–4)**; cognitive reappraisal (46.89, 1.0E–4); expressive suppression (31.32, 1.0E–4); executive function (24.03, 1.0E–4); path analysis (18.02, 1.0E–4) |
| 12 | 14 | 0.937 | 2017 | **mild cognitive impairment (17.46, 1.0E–4)**; tea consumption (8.72, 0.005); writing performance (8.72, 0.005); early childhood education (8.72, 0.005); metacomprehension assessment (8.72, 0.005) |
| 13 | 8 | 0.949 | 2021 | **virtual reality (28.75, 1.0E–4)**; grief (28.75, 1.0E–4); bereavement (22.05, 1.0E–4); phenomenological study (9.55, 0.005); sense of self (9.55, 0.005) |

## 5. Discussion

This study is based on the journal *Current Psychology*, visualizing and analyzing its annual publications, citation frequencies, countries, institutions, and keywords. The research aims to explore the journal's current research hotspots and emerging trends. A total of 4,869 articles from *Current Psychology* were retrieved from WOS. The most prolific author was Mark D. Griffiths, with 17 publications. Beijing Normal University, Southwest University, and Renmin University of China emerged as the leading institutions in terms of publications, with 176, 70, and 61 articles respectively. There was active collaboration among authors, countries, and institutions, with research hotspots in *Current Psychology* focusing on work engagement, young people, mental health, parenting stress, and related areas. Institutions and scholars worldwide are dedicated to investigating these hotspots and actively pursuing strategies to address associated issues, reflecting the developmental trends of the journal.

Based on the data presented above, we can preliminarily paint an overall picture of *Current Psychology* over the past decade. Throughout its development, *Current Psychology* has demonstrated a qualitative increase in annual publications, citations, and other metrics, illustrating rapid progress in its field. However, such a significant expansion of the journal may potentially impact its impact factor, which stands at 2.5, indicating a decline compared to the previous two years. *Current Psychology* welcomes substantial and rigorous empirical and theoretical contributions across all fields of psychology. By fostering sustained innovation and cross-disciplinary integration, the journal has steadily bolstered its visibility and influence within relevant academic domains. In recent years, *Current Psychology* has expanded its publications significantly. However, it should concurrently enhance its scrutiny of article quality to maintain high publication standards while focusing on improving its impact factor. It should not prioritize increasing publications and citations at the expense of its impact factor. Additionally, analysis of co-authors and co-cited references analysis revealed that works originate not only from prolific authors. Therefore, developing countries must encourage institutional participation in research, foster collaboration, promote advancements in relevant fields, and publish high-quality articles.

Based on the bibliometric analysis of *Current Psychology*'s publications over the past decade, this paper outlines the anticipated development trends for the journal.

Firstly, this paper predicts that publications and citations in *Current Psychology* will continue to increase. On the one hand, as academic journals gain increasing prominence, high-quality journals like *Current Psychology* are becoming more popular among domestic researchers. This expansion of the journal's submission scope and quantity is positive news for domestic researchers. On the other hand, many journals with limited influence and academic standing may be more inclined to publish low-quality articles to sustain their operation and financial viability. While an increase in the number of publications by a journal can enhance its visibility and influence, it primarily hinges on the journal's ability to maintain rigorous quality control. Inadequate quality control can potentially lead the journal into a trap of publishing low-quality articles.

Hence, there is a need to promote interdisciplinary research to study social psychological issues using diverse research methods, as real-world problems are not confined to a single discipline. Psychology is an open and pivotal interdisciplinary discipline (Borghi & Fini, 2019), and it has increasingly become a focal area of research for researchers who continuously absorb and integrate knowledge from diverse disciplines. Cross-disciplinarity is now a major trend, emphasizing the need and importance of strengthening psychology's integration with diverse fields such as medicine, education, and management. Multidisciplinary team research integrates methods and theories from various disciplines to achieve common goals (Proctor & Vu, 2019). For instance, enhancing the utilization of large-scale panel data in educational psychology and employing specialized models to explore various combinations of covariates (Kiefer et al., 2023). Embracing advanced techniques and research methods to substantiate psychological phenomena, thereby advancing psychology from theoretical discourse to empirical evidence.

Secondly, this paper emphasizes the importance of maintaining the journal's quality and suggests improvements while preserving its current impact factor. The decrease in the impact factor of *Current Psychology* in recent years can be attributed to the journal's expansion. To maintain the quality, reputation, and influence of the journal, it is crucial to uphold high article standards and refrain from publishing articles of inconsistent quality. High volumes of publications of varying quality will ultimately harm the journal and are not conducive to its long-term healthy and stable development. To achieve genuine progress, all aspects of the journal must advance in harmony, ensuring continuous improvement. Therefore, we will delve into current hot topics in contemporary psychology and explore aspects of personality psychology in adolescents and adults. *Current Psychology* should address the actual needs of the Party and the State, focusing on studying psychological issues among diverse groups in the current era, thereby deepening its research orientation centered on real-world problems. Enhancement of the contextualization of personality asserts that any personality must rely on interaction with the situation (Baumeister, 1999; Funder, 2006). This synergy holds potential for unique scientific contributions in psychology, as demonstrated by authors of this special issue (Roberts, 2007).

Thirdly, this paper suggests that *Current Psychology* should address various psychological issues from multiple perspectives, particularly focusing on topics related to the physical and mental health development of individuals and the social pressures they bear. The journal should continue to focus on maintaining its distinctive features. This will facilitate more effective research on stress, emotions, and other focal issues, leading to more precise solutions and enabling research findings to become a productive force in problem-solving. These issues are particularly relevant in today's fast-paced, high-pressure modern society and hold significant research value. Integrated analysis of the bibliometric data can aid in finding solutions to these problems and serve humanity. To strengthen the journal's identity, it is crucial to emphasize the emotional psychology of individuals and the interaction with social reality from a people-oriented perspective. With the passage of time, psychology has increasingly focused on the interaction between individuals and all aspects of social reality. The study of emotional psychology has thus become even more pertinent to real-life applications. At present, the study of emotions and emotion-related categories, which link organism variables with social environment variables, has become a focal point in current psychological research (Jia et al., 2019). For example, researchers can study the emotional changes among adolescents and the factors influencing them in the post-epidemic era. It is also essential to enhance analysis and research on the negative emotions experienced by groups due to the current era (Zhang et al., 2020). Due to the significantly prevalent psychological distress caused by COVID-19, issues may persist even after the pandemic ends, particularly affecting groups with heightened vulnerability, such as young adult females, individuals exposed to COVID-19 within their immediate social network, and those who experienced viral symptoms during the outbreak (Kibbey et al., 2021).

In terms of theoretical contribution, this paper provides valuable references for subsequent researchers. Through bibliometric analysis and visualization tools, it conducts co-citation and co-occurrence analyses of keywords to quantitatively and objectively analyze related fields, revealing the quantitative relationships among various studies. In terms of practical value, after conducting visualization analysis, this study allows for a grasp of the current hotspots of the journal and foresight into future development trends. This understanding helps elucidate broader trends in psychology, facilitating the commercialization of scientific research results. It also provides solutions to current social issues, aids in resolving psychological problems, alleviates stress, and mitigates social conflicts.

**Data availability** The data are available from the corresponding author upon reasonable request.

## Declarations

**Competing interests** The authors declare that the research was conducted in the absence of any commercial or financial relationships that could be construed as a potential conflict of interest.

# Supplementary Materials

**Table S1 Top 20 highly cited authors in *Current Psychology***

| Rank | Authors | Citations | Year |
|---|---|---|---|
| 1 | Li Zhang | 32 | 2019 |
| 2 | Ye Wang | 19 | 2021 |
| 3 | Ye Li | 18 | 2021 |
| 4 | Yan Zhang | 17 | 2022 |
| 5 | Jing Wang | 16 | 2021 |
| 6 | Li Lei | 15 | 2020 |
| 7 | Li Wang | 15 | 2021 |
| 8 | Hui Chen | 14 | 2022 |
| 9 | Mark D. Griffiths | 14 | 2022 |
| 10 | Jing Li | 14 | 2021 |
| 11 | Xue Li | 14 | 2022 |
| 12 | Xu Wang | 14 | 2022 |
| 13 | Yanhua Wang | 14 | 2018 |
| 14 | Yang Liu | 13 | 2022 |
| 15 | Xuqun You | 13 | 2017 |
| 16 | Qi Zhang | 13 | 2022 |
| 17 | Julia Brailovskaia | 12 | 2020 |
| 18 | Claudia Ferreira | 12 | 2020 |
| 19 | Xinyi Li | 12 | 2021 |
| 20 | Jürgen Margraf | 12 | 2020 |

**Table S2 Top 20 institutions in *Current Psychology* by publications**

| Rank | Institutions | Publications | Counries |
|---|---|---|---|
| 1 | Beijing Normal University | 176 | People's Republic of China |
| 2 | Southwest University | 70 | People's Republic of China |
| 3 | Renmin University of China | 61 | People's Republic of China |
| 4 | Ohio University | 57 | The United States |
| 5 | Central China Normal University | 56 | People's Republic of China |
| 6 | Chinese Academy of Sciences | 54 | People's Republic of China |
| 7 | Shaanxi Normal University | 51 | People's Republic of China |
| 8 | Sapienza University of Rome | 50 | Repubblica Italiana |
| 9 | Universidade de Coimbra | 49 | The Portuguese Republic |
| 10 | University of London | 46 | The United Kingdom |
| 11 | South China Normal University | 43 | People’s Republic of China |
| 12 | Universitat de València | 41 | The Kingdom of Spain |
| 13 | Zhejiang University | 41 | People’s Republic of China |
| 14 | University of Macau | 40 | People’s Republic of China |
| 15 | East China Normal University | 38 | People’s Republic of China |
| 16 | University of Warsaw | 38 | The Republic of Poland |
| 17 | The Education University of Hong Kong | 35 | People’s Republic of China |
| 18 | Middle East Technical University | 33 | The Republic of Türkiye |
| 19 | Shanghai Normal University | 33 | People’s Republic of China |
| 20 | University of California | 33 | The United States |

**Table S3 Top 20 countries in *Current Psychology* by publications**

| Rank | Countries | Publications | Year |
|---|---|---|---|
| 1 | People’s Republic of China | 1355 | 2014 |
| 2 | The United States | 933 | 2013 |
| 3 | The Republic of Türkiye | 419 | 2014 |
| 4 | The United Kingdom | 284 | 2013 |
| 5 | The Kingdom of Spain | 273 | 2013 |
| 6 | Repubblica Italiana | 241 | 2014 |
| 7 | Federal Republic of Germany | 240 | 2013 |
| 8 | Canada | 219 | 2014 |
| 9 | The Republic of Poland | 187 | 2013 |
| 10 | The Islamic Republic of Iran | 186 | 2014 |
| 11 | The Commonwealth of Australia | 176 | 2013 |
| 12 | The Portuguese Republic | 123 | 2015 |
| 13 | Republic of Korea | 123 | 2014 |
| 14 | Japan | 108 | 2015 |
| 15 | The Kingdom of the Netherlands | 100 | 2013 |
| 16 | The State of Israel | 93 | 2013 |
| 17 | The Republic of India | 83 | 2017 |
| 18 | The French Republic | 80 | 2014 |
| 19 | Malaysia | 79 | 2016 |
| 20 | Islamic Republic of Pakistan | 71 | 2015 |

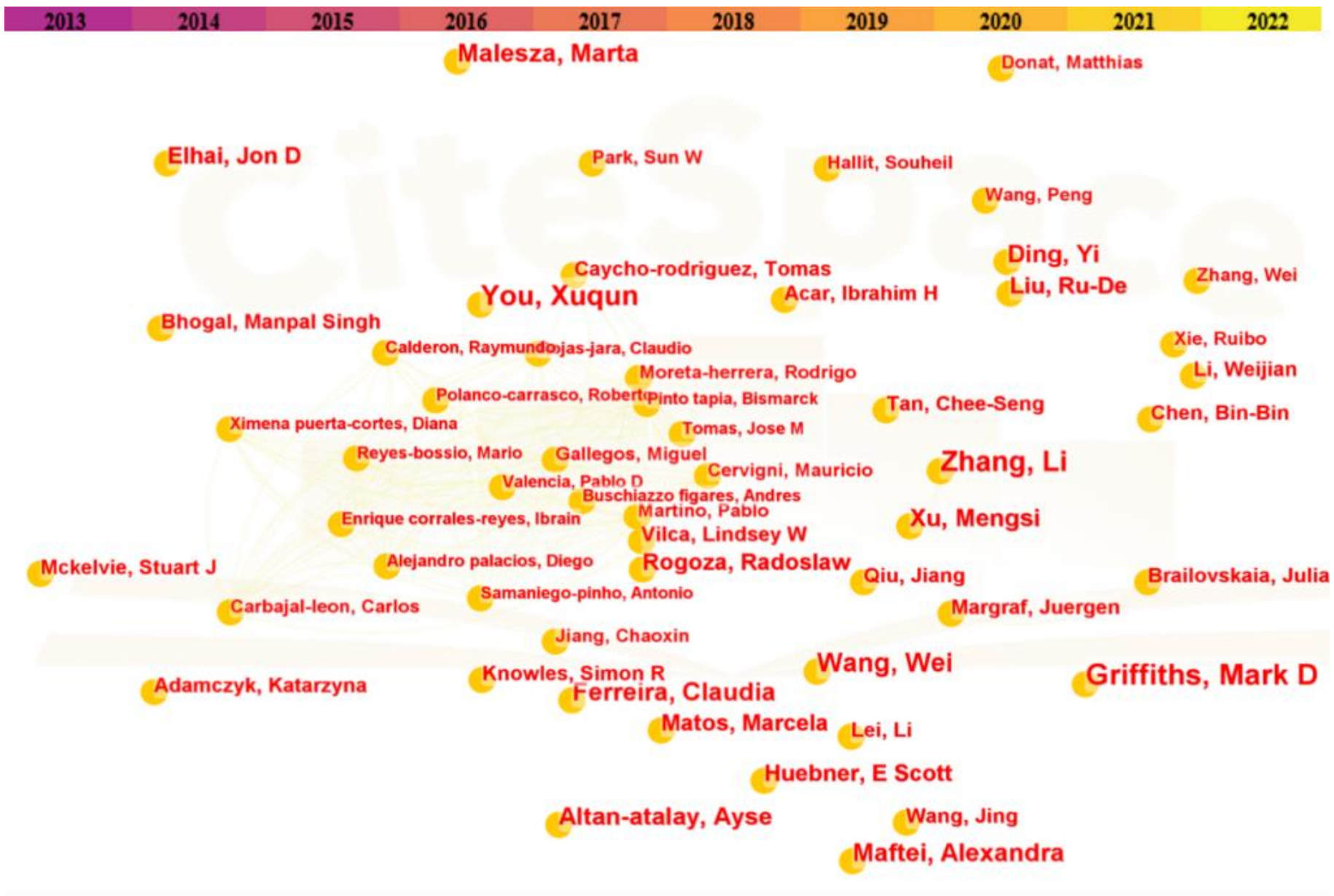

**Figure S1. Co-occurrence network of authors**

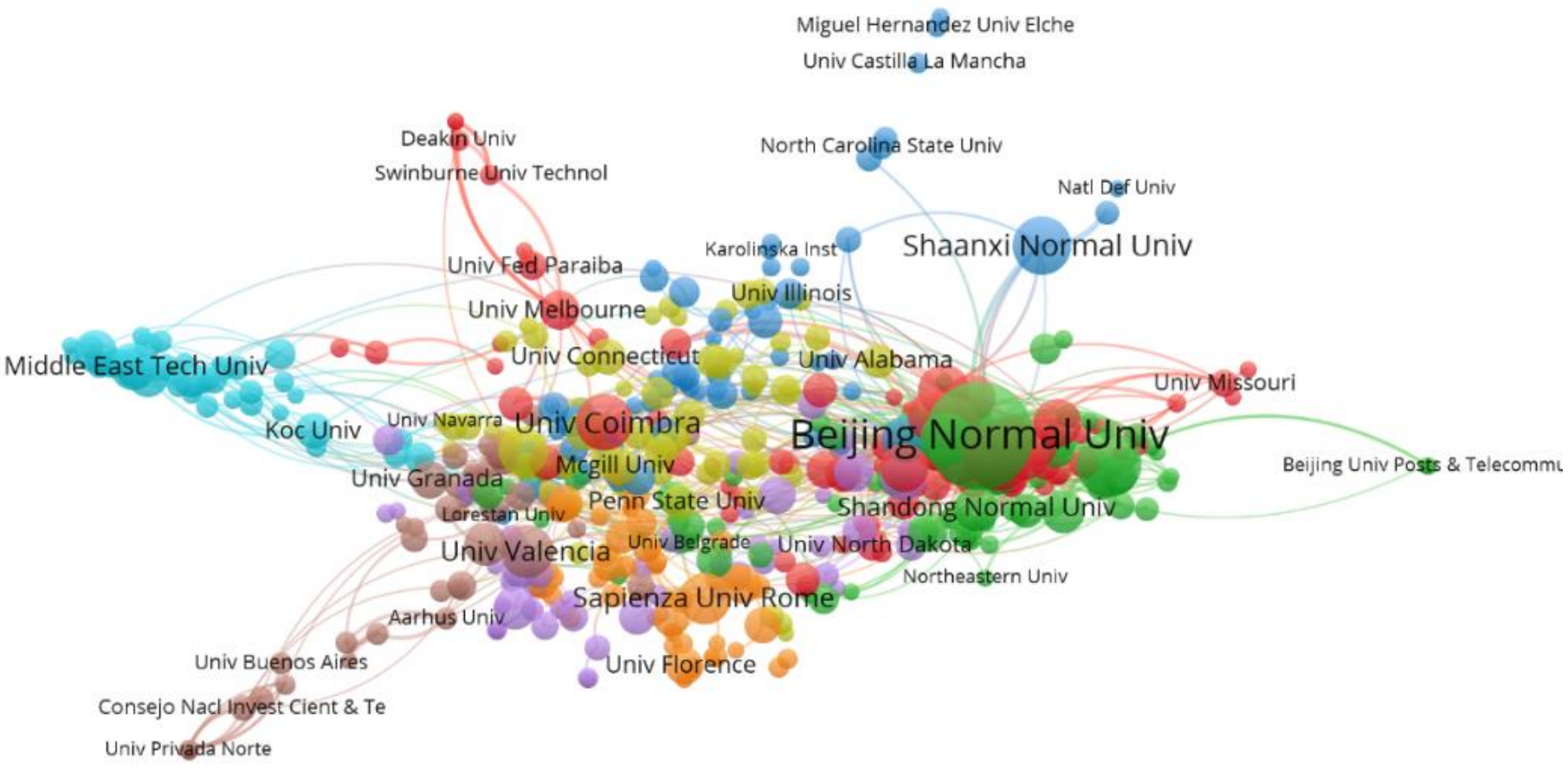

**Figure S2. Co-occurrence network of insitituions**

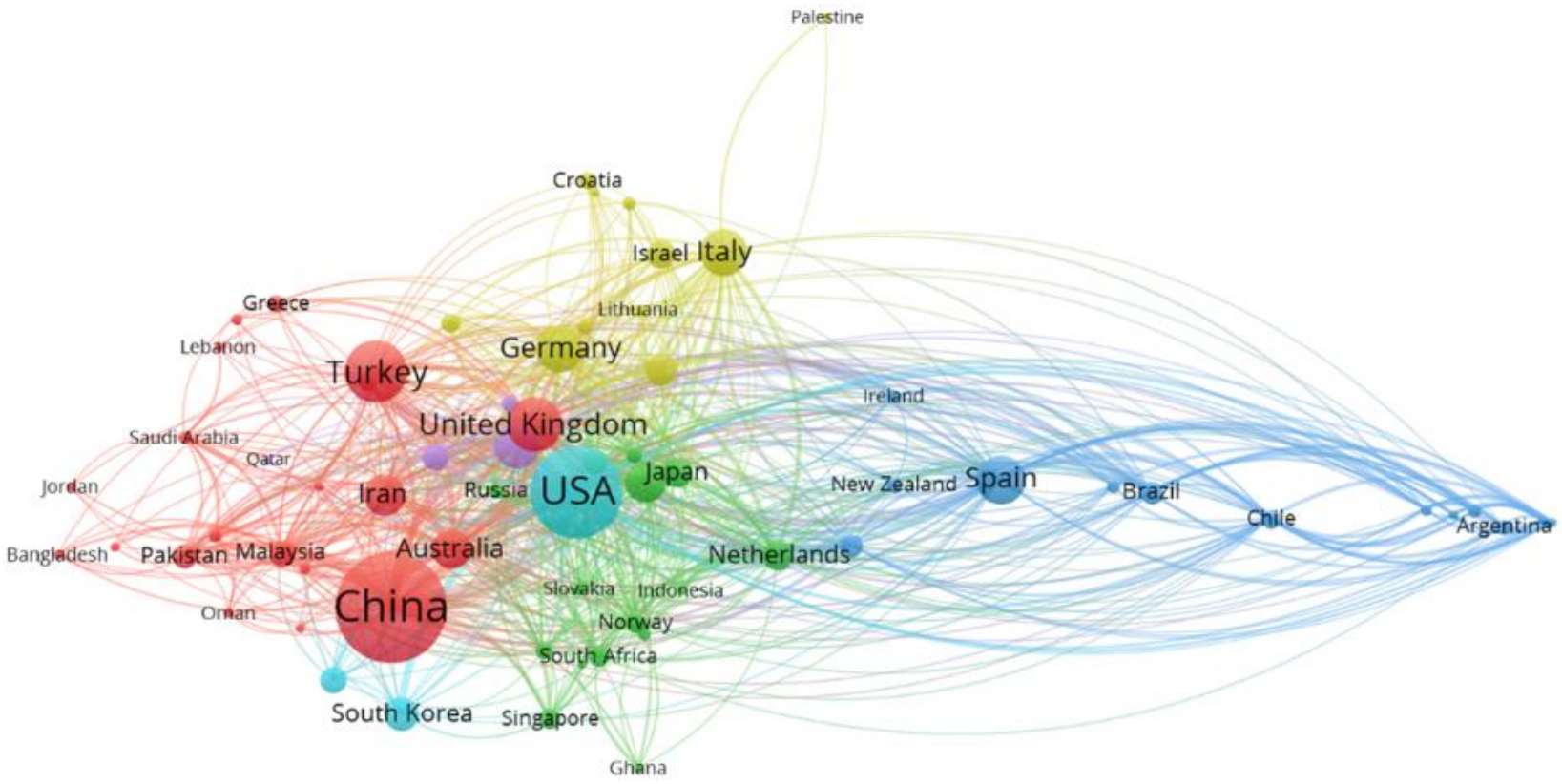

**Figure S3. Countries' clusters view**

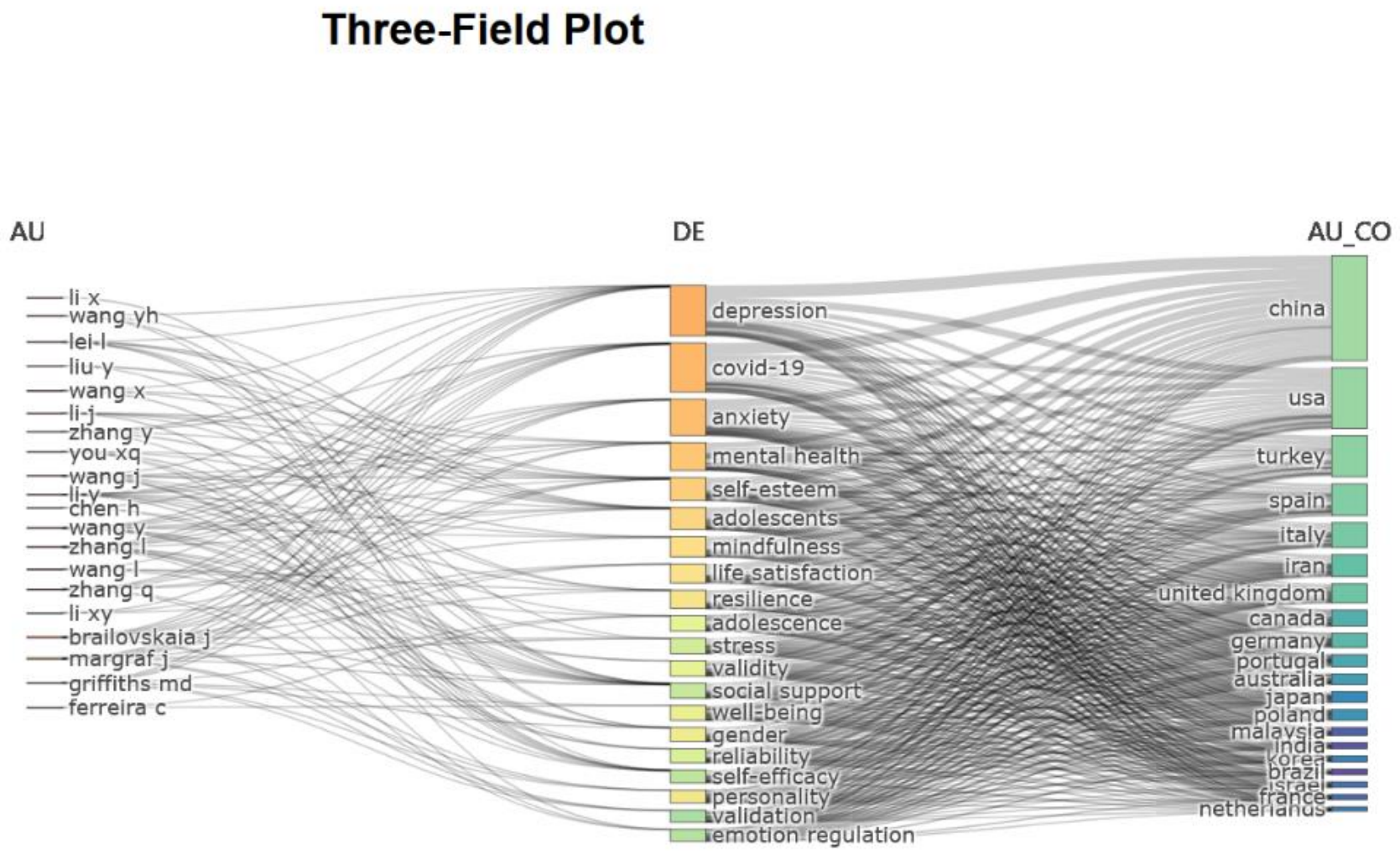

**Figure S4 Mulberry Figure of *Current Psychology* keywords connected author and country**

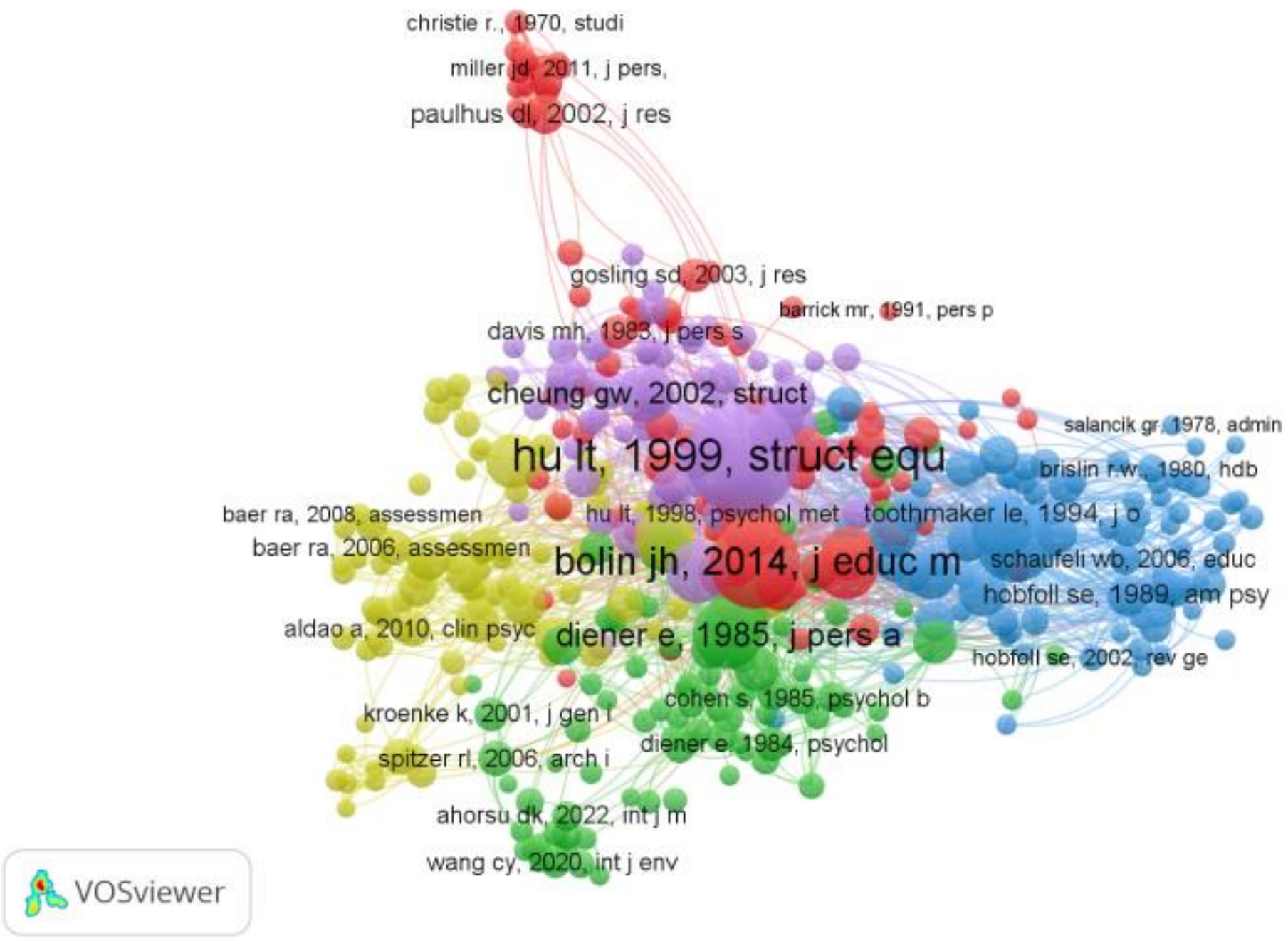

**Figure S5. References' clusters view**

**Top 20 References with the Strongest Citation Bursts**

| References | Year | Strength | Begin | End | 2013 - 2022 |
|---|---|---|---|---|---|
| American Psychiatric Association, 2013, DIAGN STAT MAN MENT, V5th, P0, DOI 10.1176/APPI.BOOKS.9780890425596, DOI | 2013 | 2.44 | **2013** | 2017 | |
| American Psychiatric Association, 2013, DIAGN STAT MAN MENT, V5th, P0, DOI 10.1176/appi.books.9780890425596, DOI | 2013 | 6.59 | **2014** | 2018 | |
| American Psychiatric Association, 2022, DIAGN STAT MAN MENT, V5, P0, DOI 10.1176/appi.books.9780890425787, 10.1176/APPI.BOOKS.9780890425787, 10.1176/APPI.BOOKS.9780890425787, DOI | 2022 | 2.55 | **2022** | 2020 | |
| Adamczyk K, 2013, REV INT PSYCHOL SOC, V26, P25 | 2013 | 1.3 | **2015** | 2016 | |
| Kline R. B., 2015, METHODOLOGY SOCIAL S, V4th, P0 | 2015 | 6.95 | **2016** | 2020 | |
| Brown T. A., 2015, CONFIRMATORY FACTOR, V0, P0, DOI 10.1007/s11336-016-9548-3, DOI | 2015 | 6.41 | **2016** | 2020 | |
| Datu JAD, 2015, CURR PSYCHOL, V34, P626, DOI 10.1007/s12144-014-9275-9, DOI | 2015 | 2.6 | **2016** | 2017 | |
| Grijalva E, 2015, PSYCHOL BULL, V141, P261, DOI 10.1037/a0038231, DOI | 2015 | 2.13 | **2016** | 2020 | |
| Adamczyk K., 2014, PSYCHOL TOPICS, V23, P327 | 2014 | 1.3 | **2016** | 2017 | |
| Hayes A. F, 2022, INTRO MEDIATION MODE, V3rd, P0 | 2022 | 18.04 | **2022** | 2020 | |
| He F, 2014, SOC INDIC RES, V119, P897, DOI 10.1007/s11205-013-0519-0, DOI | 2014 | 3.91 | **2017** | 2018 | |
| Hayes A. F., 2013, INTRO MEDIATION MODE, V0, P0 | 2013 | 3.26 | **2017** | 2018 | |
| Alt D, 2018, CURR PSYCHOL, V37, P128, DOI 10.1007/s12144-016-9496-1, DOI | 2018 | 2.22 | **2018** | 2020 | |
| Acar IH, 2015, EARLY EDUC DEV, V26, P479, DOI 10.1080/10409289.2015.1000718, DOI | 2015 | 1.73 | **2018** | 2020 | |
| Muthen L.K., 2017, MPLUS CHAPMAN HALL, V8th ed., P0 | 2017 | 2.23 | **2019** | 2020 | |
| Krizan Z, 2018, PERS SOC PSYCHOL REV, V22, P3, DOI 10.1177/1088868316685018, DOI | 2018 | 3.02 | **2020** | 2022 | |
| Elhai JD, 2016, COMPUT HUM BEHAV, V63, P509, DOI 10.1016/j.chb.2016.05.079, DOI | 2016 | 2.69 | **2020** | 2022 | |
| Hodson G, 2018, J RES PERS, V73, P123, DOI 10.1016/j.jrp.2017.11.012, DOI | 2018 | 2.35 | **2020** | 2022 | |
| Bleidorn W, 2016, J PERS SOC PSYCHOL, V111, P396, DOI 10.1037/pspp0000078, DOI | 2016 | 2.01 | **2020** | 2022 | |
| Beyens I, 2016, COMPUT HUM BEHAV, V64, P1, DOI 10.1016/j.chb.2016.05.083, DOI | 2016 | 2.01 | **2020** | 2022 | |

**Figure S6. Burst view of references**

**Top 25 Keywords with the Strongest Citation Bursts**

| Keywords | Year | Strength | Begin | End | 2013 - 2022 |
|---|---|---|---|---|---|
| gender differences | 2013 | 8.84 | **2013** | 2019 | |
| performance | 2013 | 6.93 | **2013** | 2018 | |
| knowledge | 2013 | 5.3 | **2013** | 2016 | |
| sex differences | 2013 | 5.2 | **2013** | 2020 | |
| perception | 2013 | 4.22 | **2013** | 2015 | |
| perceptions | 2013 | 3.69 | **2013** | 2019 | |
| attitudes | 2014 | 10.59 | **2014** | 2018 | |
| memory | 2014 | 7.17 | **2014** | 2018 | |
| social support | 2014 | 6.83 | **2014** | 2019 | |
| behavior | 2014 | 5.89 | **2014** | 2018 | |
| physical attractiveness | 2014 | 4.14 | **2014** | 2019 | |
| self | 2015 | 9.14 | **2015** | 2018 | |
| personality | 2014 | 15.54 | **2016** | 2020 | |
| scale | 2014 | 9.37 | **2016** | 2019 | |
| satisfaction | 2014 | 6.42 | **2016** | 2018 | |
| ideology | 2016 | 4.43 | **2016** | 2020 | |
| college students | 2014 | 6.01 | **2017** | 2020 | |
| adult attachment | 2013 | 3.78 | **2017** | 2020 | |
| self regulation | 2018 | 4.55 | **2018** | 2020 | |
| values | 2018 | 3.85 | **2018** | 2020 | |
| commitment | 2015 | 3.72 | **2018** | 2019 | |
| emotional intelligence | 2018 | 3.66 | **2018** | 2019 | |
| individual differences | 2016 | 10.26 | **2019** | 2020 | |
| reliability | 2014 | 6.59 | **2019** | 2020 | |
| dark triad | 2019 | 3.71 | **2019** | 2020 | |

**Figure S7. Burst view of top 25 keywords**

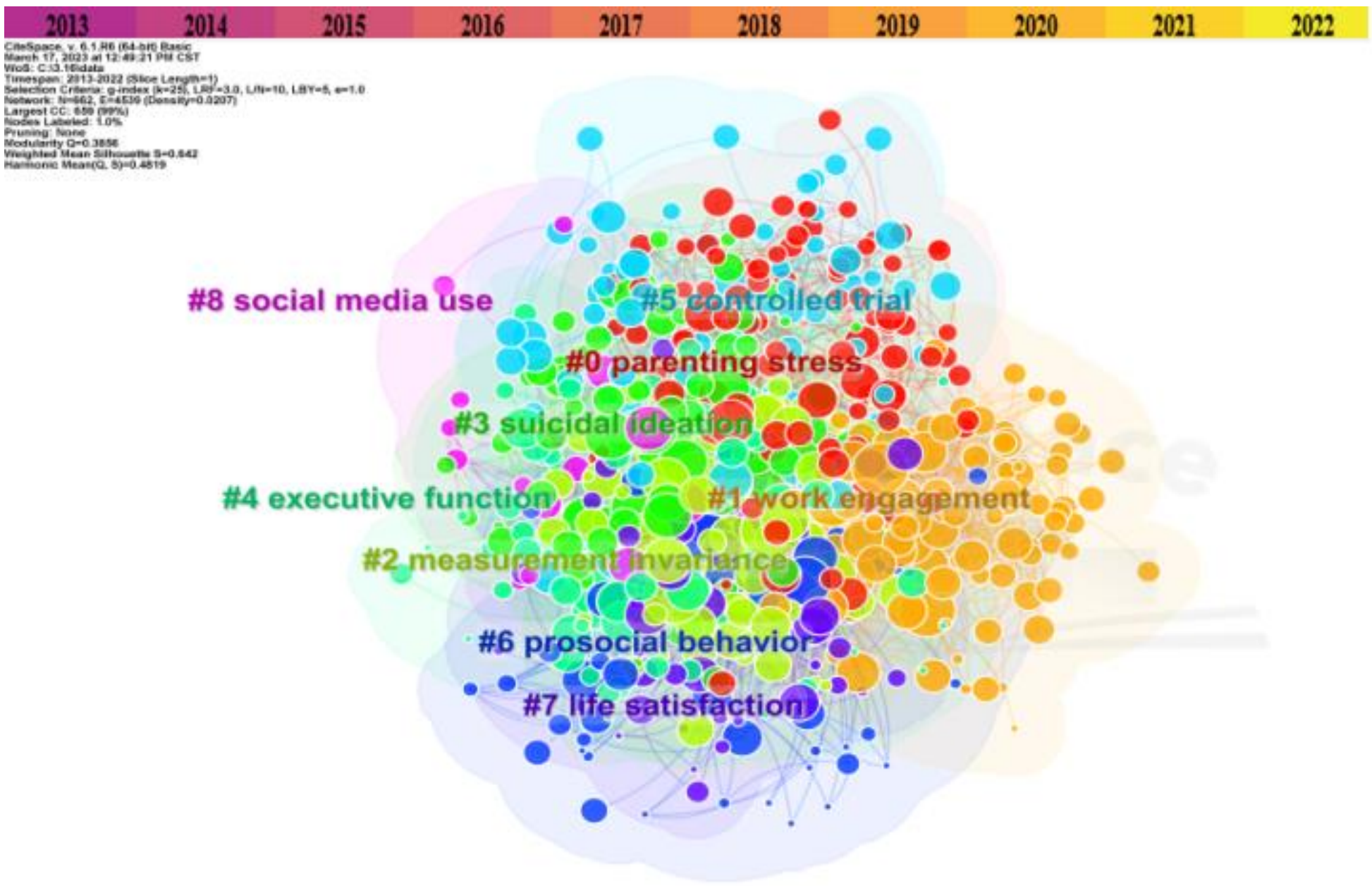

**Figure S8. Thematic keywords cluster network via CiteSpace**